\documentclass[letter]{aa}  

\usepackage{graphicx}
\usepackage{gensymb}
\usepackage{txfonts}
\begin{document} 
   
   \title{A diffuse tidal dwarf galaxy destined to fade out as a "dark galaxy"}

   \author{Javier Román\inst{1,2,3}\thanks{   \email{jromanastro@gmail.com}}
   \and
   Michael G. Jones\inst{1,4}
   \and
   Mireia Montes\inst{5,6}\thanks{STScI Prize Fellow}
   \and
   Lourdes Verdes-Montenegro\inst{1}
   \and
   Julián Garrido\inst{1}
   \and
   Susana Sánchez\inst{1}
   }

   \institute{Instituto de Astrof\'isica de Andaluc\'ia (CSIC), Glorieta de la Astronom\'ia, 18008 Granada, Spain
   \and Instituto de Astrof\'{\i}sica de Canarias, c/ V\'{\i}a L\'actea s/n, E-38205, La Laguna, Tenerife, Spain
   \and Departamento de Astrof\'{\i}sica, Universidad de La Laguna, E-38206, La Laguna, Tenerife, Spain
   \and Steward Observatory, University of Arizona, 933 North Cherry Avenue, Rm. N204, Tucson, AZ 85721-0065, USA
   \and Space Telescope Science Institute, 3700 San Martin Drive, Baltimore, MD 21218, USA
   \and School of Physics, University of New South Wales, Sydney, NSW 2052, Australia
}
   
   \date{\today}

 
  \abstract
   {We have explored the properties of a peculiar object detected in deep optical imaging and located at the tip of an H\,{\sc i} tail emerging from Hickson~Compact~Group~16. Using multiband photometry from infrared to ultraviolet, we were able to constrain its stellar age to 58$^{+22}_{-9}$~Myr with a rather high metallicity of [Fe/H] = $-$0.16$^{+0.43}_{-0.41}$ for its stellar mass of M$_\star$~=~4.2$\times$10$^6$~M$_\odot$, a typical signature of tidal dwarf galaxies. The structural properties of this object are similar to those of diffuse galaxies, with a round and featureless morphology, a large effective radius (r$_{eff}$ = 1.5 kpc), and a low surface brightness (<$\mu_{g}$>$_{eff}$ = 25.6 mag arcsec$^{-2}$). Assuming that the object is dynamically stable and able to survive in the future, its fading in time via the aging of its stellar component will make it undetectable in optical observations in just $\sim$2 Gyr of evolution, even in the deepest current or future optical surveys. Its high H\,{\sc i} mass, M(HI)~=~3.9$\times$10$^8$~M$_\odot$, and future undetectable stellar component will make the object match the observational properties of dark galaxies, that is, dark matter halos that failed to turn gas into stars. Our work presents further observational evidence of the feasibility of H\,{\sc i} tidal features becoming fake dark galaxies; it also shows the impact of stellar fading, particularly in high metallicity systems such as tidal dwarfs, in hiding aged stellar components beyond detection limits in optical observations.
   }


   \maketitle
%

\section{Introduction}

Our understanding of the Universe is closely tied to our capabilities to observe it, and surface brightness limits create severe biases that must be taken into account in order to provide a correct representation of the low-mass galactic population \citep{1976Natur.263..573D, 1995AJ....110..573M}. While the new generation {of} deep optical surveys are providing a growing number of previously undetected low surface brightness galaxies \citep[e.g.,][]{2018ApJ...857..104G, 2019ApJS..240....1Z, 2020arXiv200604294T}, the surface brightnesses achieved by integrated photometry are far from the limits achieved by star counting in the local Universe, where galaxies of up to 32 mag arcsec$^{-2}$ have been discovered \citep{2019PASJ...71...94H}. There is, therefore, a huge population of currently undetectable low-mass galaxies beyond the distance at which we can resolve individual stars.

An example of the impact of surface brightness limitations is found in the search for the so-called dark galaxies, that is, dark matter halos that failed to convert their gas into stars. They are expected from a theoretical perspective because of the low star formation efficiency of low mass halos at low metallicity \citep{2010ApJ...714..287G, 2012ApJ...753...16K}. Objects compatible with dark galaxies have been detected at high redshifts in Ly$\alpha$ fluorescence by being illuminated by quasars \citep{2012MNRAS.425.1992C, 2018ApJ...859...53M, 2019ApJ...875..130L}. However, the systematic search for dark galaxies in the local Universe is strongly biased by the prevalence of H\,{\sc i} debris expelled from galactic interactions that can masquerade as gas within a dark matter halo that failed to form stars, {that is, a genuine dark galaxy} \citep[see][]{2005MNRAS.363L..21B, 2008ApJ...673..787D}. In fact, H\,{\sc i} clouds with no optical counterpart are common in high-density environments \citep[e.g.,][]{2012MNRAS.423..787T,2013A&A...555L...7O,2017MNRAS.464..530S,2020A&A...642L..10B}, but systematic searches in low-density environments (where the possibility of H\,{\sc i} tidal debris can be minimized) have not returned any clear candidates for true dark galaxies \citep{2015AJ....149...72C}, calling into question the existence of bona fide dark galaxies. 

{Gas-rich} tidal interactions are commonly accompanied by bursts of star formation \citep{2001ApJ...550..204I, 2005A&A...430..443V, 2012MNRAS.426.2441D, 2014MNRAS.443.3601L} due to local instabilities of the gas \citep{2007MNRAS.375..805W}, forming the so-called tidal dwarf galaxies \citep[TDGs; ][]{1998A&A...333..813D}. Due to the difficulty in obtaining {a} high resolution surface density of the gas needed to establish star formation thresholds in these systems \citep{2004ApJ...609..667S}, the potential relation between tidal interactions and H\,{\sc i} clouds without an optical component is hard to study observationally. Additionally, effects such as fading luminosity due to the aging of stellar populations have to be properly taken into account in order to assert the presence or absence of a stellar component.

\begin{figure*}
    \centering
        \includegraphics[width=0.97\textwidth]{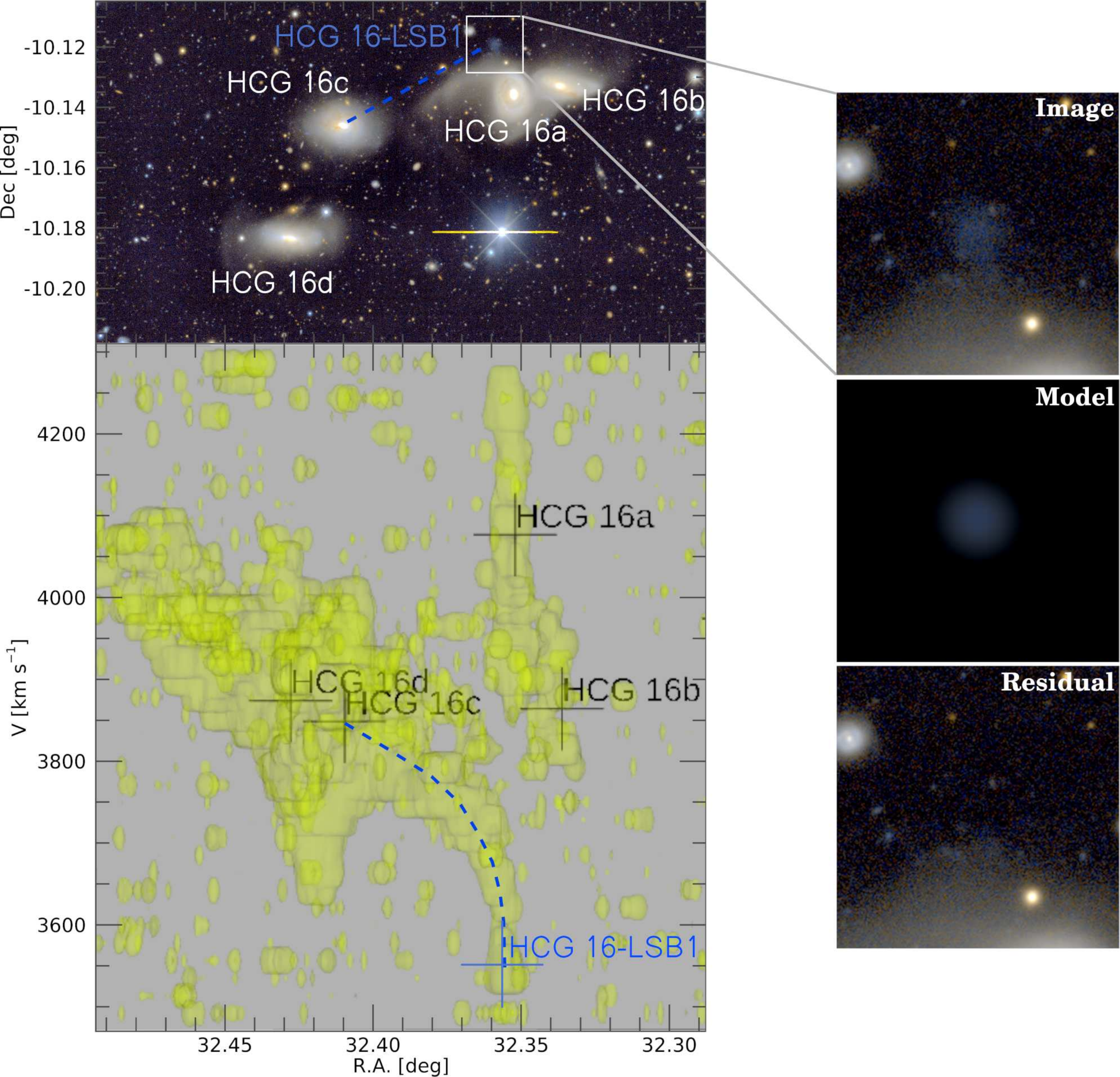}
    \caption{Illustration of the structures present in the northwestern region of HCG16. Upper left panel: Color-composed image using the \textit{g, r}, and \textit{z} optical bands from DECaLS. Lower left panel: Pseudo zenith view (R.A. vs. H\,{\sc i} velocity) of the H\,{\sc i} emission higher than 3$\sigma_\mathrm{rms}$. {Integrated moment-zero and first moment maps as well as detailed information about these data and maps for the whole group can be found in \cite{2019A&A...632A..78J}}. The dashed blue line in both left panels shows the trajectory of the H\,{\sc i} tail, the end of which coincides with the diffuse object detected in the optical. {We note that the progenitor of HCG 16-LSB1 is most likely HCG 16c and not the apparently closest galaxy (in projection), HCG 16a}. Right panels: Zoom-in of the object and its fitting using optical bands through a S\'ersic model.}
    \label{fig:Panel}
\end{figure*}

In this letter we study the properties of a peculiar object that was identified by \cite{2019A&A...632A..78J} as a very faint feature detected in optical data {and} located at the end of an H\,{\sc i} tail emerging from the complex interactions of Hickson Compact Group 16 (HCG 16), which \cite{2019A&A...632A..78J} suggest is a clear TDG candidate. For convenience, we have named this object HCG 16-LSB1. We aim to test its TDG nature, study its properties, and analyze its potential future evolution. We assume standard cosmology parameters, with a distance to the object of 54.7 Mpc and a spatial scale of 0.259 kpc arcsec$^{-1}$. We use the AB photometric magnitude system in this work.

\section{Properties of HCG 16-LSB1}

We used publicly available photometric data from the seventh {data} release of the Dark Energy Camera Legacy Survey  \citep[DECaLS; ][]{2016MNRAS.460.1270D} to obtain the photometric and structural properties of HCG 16-LSB1 in the optical. We performed a S\'ersic model fitting using \texttt{IMFIT} \citep{2015ApJ...799..226E} in the three available photometric bands (\textit{g}, \textit{r}, and \textit{z}). The parameters resulting from this modeling are shown in Table \ref{tab:phot}. These properties match with those of diffuse galaxies frequently found in groups of galaxies \citep[e.g.,~][]{2017MNRAS.468.4039R}, with $r_{e}$~=~1.5~kpc, $\mu(0)_{g}$~=~25.2~mag~arcsec$^{-2}$, and $n$~=~0.48. The colors of HCG 16-LSB1 are extremely blue, with \textit{g-r}~=~0.00~$\pm$~0.10 mag, suggesting very young stellar populations. In the right panels of Fig. \ref{fig:Panel} we plot the color composite model of the object together with its residuals. As can be seen, a S\'ersic model is a good representation of the object with a clean residual image. 

We further explored other optical data sets. We found observations with the Canada-France-Hawaii Telescope (CFHT) in the CFHT Science Archive for the \textit{g}, \textit{r}, and \textit{i} optical bands. However, the presence of an internal reflection from a nearby star prevents photometry of the object in these images. We also found observations with the Hubble Space Telescope (HST) in the Hubble Legacy Archive (PropID = 10787) for the \textit{F814W}, \textit{F606W}, and \textit{F450W} bands. In this case, we ruled out the use of these images due to the presence of residuals in the reduction process that have a strong impact on low surface brightness objects in the HST images; these images would thus require complex data processing in order to obtain reliability \citep{2019A&A...621A.133B}. While photometric or structural measurements in these additional optical data are unfeasible, their high resolution allows us to explore in great detail the possible presence of HII regions. However, we do not find any such regions within the depth and resolution limits that these data provide: HCG 16-LSB1 appears to be a smooth and featureless galaxy.

\subsection{SED fitting}

We conducted a study of the stellar populations of HCG 16-LSB1 with the available photometric information in this region, aiming to constrain its age and metallicity. In addition to the aforementioned data in the \textit{g}, \textit{r}, and \textit{z} optical bands, we also added the ultraviolet data from the Galaxy Evolution Explorer \citep[GALEX;][]{2005ApJ...619L...1M} and infrared archival data from Spitzer \citep{2004ApJS..154....1W}. The photometry was carried out by fitting a S\'ersic model, in a similar way to what was done in the optical bands. The resulting photometry is given in Table \ref{tab:phot}. In the case of Spitzer data, HCG 16-LSB1 was not detected, so upper limits were estimated.

\begin{table}
\begin{center}
\begin{tabular}{cc}
Parameter & Value \\ \hline
R.A. & 32.3587\\
Dec. & -10.1194\\
r$_e$ & 1.5 $\pm$ 0.1 kpc\\
$<\mu _g>$ & 25.6 $\pm$ 0.1 mag/arcsec$^2$ \\
$\mu _g$(0) & 25.2 $\pm$ 0.1 mag/arcsec$^2$ \\
S\'ersic index & 0.48 $\pm$ 0.01\\
Axis ratio & 0.96 $\pm$ 0.01 \\
M$_{g}$ & $-$14.05 $\pm$ 0.04 mag\\
H\,{\sc i} velocity & 3552~$\pm$~2~km~s$^{-1}$\\
\\
Far-UV band & 20.24 $\pm$ 0.45 mag\\
Near-UV band & 20.30 $\pm$ 0.15 mag\\
g-band & 19.64 $\pm$ 0.04 mag\\
r-band & 19.64 $\pm$ 0.06 mag\\
z-band & 19.94 $\pm$ 0.20 mag\\
IRAC1$_{3.6 \mu m}$ band & < 20.19 mag\\
IRAC2$_{4.5 \mu m}$ band & < 20.19 mag\\
IRAC3$_{5.8 \mu m}$ band & < 18.68 mag\\
IRAC4$_{8.0 \mu m}$ band & < 18.68 mag\\
\newline
\end{tabular}
\caption{Summary of the properties of HCG 16-LSB1. All photometric values are corrected from extinction \citep{2011ApJ...737..103S}.}
\label{tab:phot}
\end{center}
\end{table}

\begin{figure}
    \centering
        \includegraphics[width=0.8\columnwidth]{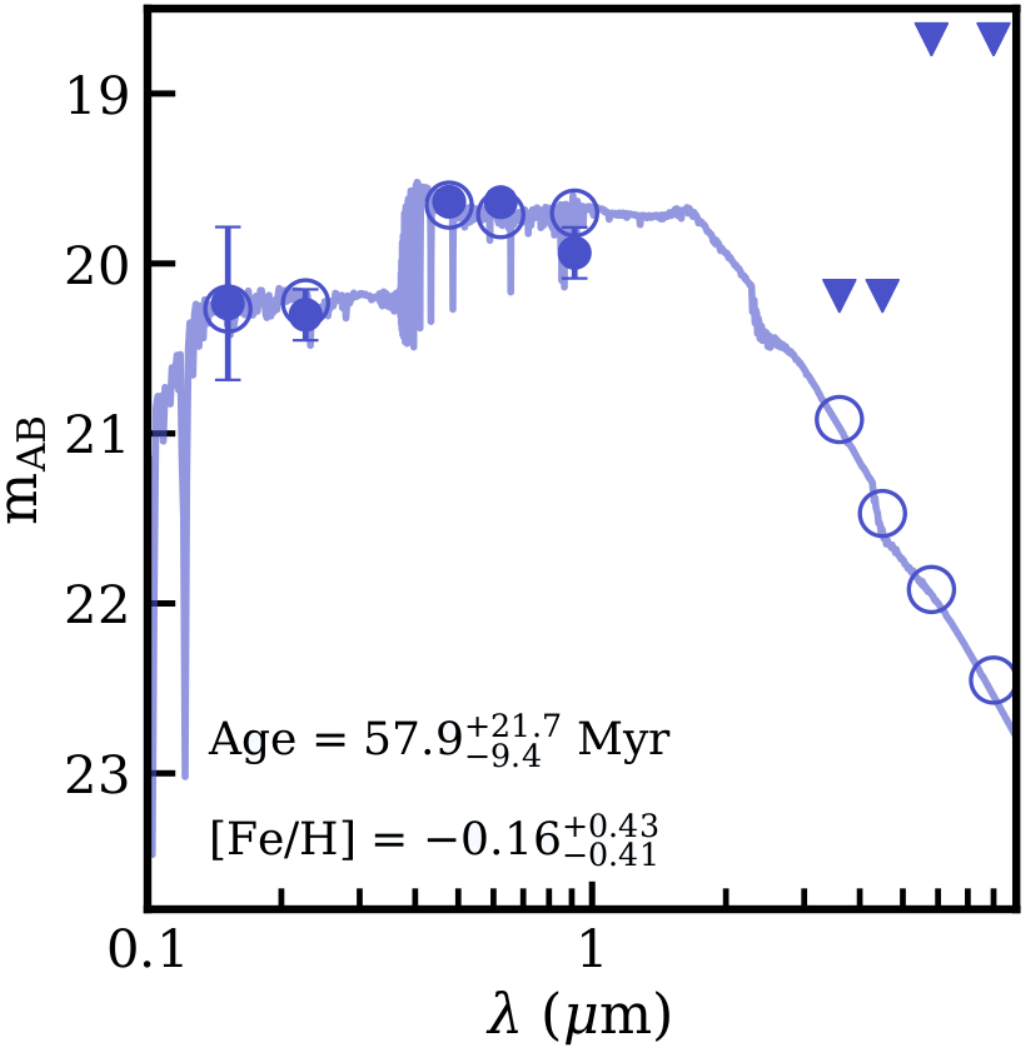}
    \caption{Spectral energy distribution of the galaxy. The filled circles are the observed magnitudes of the galaxy, while the inverted triangles correspond to the upper limits in those bands where the galaxy was not detected. The open circles are the expected flux after convolving the best fit model with the filter responses.}
    \label{fig:SED}
\end{figure}

As HCG 16-LSB1 shows no structure or "clumpiness" and has an expected low age, we assume that it formed in a single burst of star formation. Therefore, characterizing it as a single stellar population (SSP) with only a few parameters, such as metallicity, age, and initial mass function (IMF), is a good approximation. We compared the observed spectral energy distribution (SED) of HCG 16-LSB1 with SSP models \citep{2003MNRAS.344.1000B}, {using} a \citet{2003ApJ...586L.133C} IMF to obtain estimates of the age and metallicity of the galaxy. We used a $\chi^2$-minimization approach to obtain the best fit model to our data \citep[as described in][]{2014MNRAS.439..990M}. The uncertainties were estimated by marginalizing the 1D probability distribution functions. Figure \ref{fig:SED} shows the best fitting model to the SED of HCG 16-LSB1 with 57.9$^{+21.7}_{-9.4}$~Myr and [Fe/H]~=~$-$0.16$^{+0.43}_{-0.41}$ (solid blue line). All photometric points fit well with the model except for the one corresponding to the \textit{z} band. We think that this could be due to background sky fluctuations that are stronger in this band, producing a higher uncertainty due to the very low surface brightness of HCG 16-LSB1. The stellar-mass-to-light ratio is M/L$_{V-band}$~=~0.073$^{+0.097}_{-0.059}$ $\Upsilon_{\sun}$, giving a stellar mass of M$_\star$~=~4.2$^{+5.6}_{-3.4}\times$10$^6$~M$_\odot$ for HCG 16-LSB1.

The metallicity of HCG 16-LSB1 of [Fe/H]~=~$-$0.16$^{+0.43}_{-0.41}$ is much higher than what would be expected for a dwarf galaxy of its stellar mass, which is typically within the range [Fe/H]~=~[-2.0,~-1.5]  \citep{2003AJ....125.1926G, 2013ApJ...779..102K, 2017A&A...606A.115H}. This suggests that it formed from the pre-enriched gas of the galactic components of HCG 16, this high metallicity content being a fundamental characteristic defining TDGs. \citep[see a review by][]{2012ASSP...28..305D}.

\subsection{Dynamics and survival}
\label{sec:dynamic}

\begin{figure}
    \centering
        \includegraphics[width=1\columnwidth]{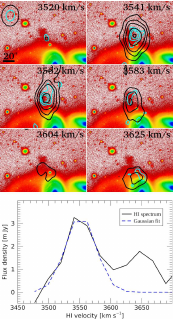}
    \caption{{H\,{\sc i} distribution of HCG 16-LSB1.} Upper panels: Velocity channel maps corresponding to high (light blue) and low (black) resolutions of the H\,{\sc i} emission plotted over a high contrast $g+r$ optical image. The beam size for both resolutions is placed in the upper left corner (see text). We note that the H\,{\sc i} emission in the plotted velocity channels is not associated with the bright galaxies (HCG 16a and HCG 16b) south of HCG 16-LSB1 (see Fig. \ref{fig:Panel}). Bottom panel: H\,{\sc i} spectrum obtained from the channels (high resolution data) presented in the upper panel (i.e., integrated over HCG 16-LSB1; see Sect. \ref{sec:dynamic} for details). The peak emission is fitted with a Gaussian function.}
    \label{fig:HI_spec}
\end{figure}

In the upper panels of Fig. \ref{fig:HI_spec} we show the H\,{\sc i} emission for different velocity channels corresponding to the tip of the H\,{\sc i} tail shown in Fig. \ref{fig:Panel}. {These data were obtained in the Very Large Array using configurations C and D and were processed by \cite{2019A&A...632A..78J}; more detailed information regarding the process is available in that work.} We plot the emission contours of the robust=0 (19.4\arcsec \ $\times$ 14.8\arcsec) and robust=2 (37.2\arcsec \ $\times$ 30.3\arcsec) cubes, with an available spectral resolution of $\approx$ 21 km s$^{-1}$. 

The velocity channels show spatial matching between the tip of the H\,{\sc i} tail and the diffuse emission in optical bands. Considering the young age and high metallicity of the stellar component, which can only be explained by a recent formation with pre-enriched material, and the fact that TDGs are preferentially formed at the tip of H\,{\sc i} tails, a chance superposition is extremely unlikely. This confirms the association between the H\,{\sc i} cloud located at the end of the H\,{\sc i} tail and the stellar component detected in optical bands. However, we find an offset between the peak emission in H\,{\sc i} and the center of the object in the optical. This offset is 5.5'' for the 3562 km s$^{-1}$ channel, which is approximately the value of the effective radius (1.5 kpc). While the association between the H\,{\sc i} tail and the stellar component seems clear, this potential shift could indicate a certain degree of gas stripping or a spatial asymmetry between both components that cannot be resolved with the resolution available in our data.

In the lower panel of Fig. \ref{fig:HI_spec} we show the integrated H\,{\sc i} spectrum. It was extracted using an aperture with dimensions twice  that of the synthesized beam of the robust=0 cube (i.e., four times the area). The robust=0 cube was favored over the robust=2 cube due to its better angular resolution; however, we note that this is at the cost of sensitivity, and thus we retain the initial H\,{\sc i} mass estimation from \cite{2019A&A...632A..78J} of M(HI)~=~3.9$\times$10$^8$~M$_\odot$, which used the robust=2 cube. We fitted the peak associated with the tip of the tail with a Gaussian function for the 3520, 3541, 3562, and 3583 km s$^{-1}$ channels, which have a spatial match with the optical counterpart. The residuals of the Gaussian fits are minimal up to 3600 km s$^{-1}$. From this velocity onward, an excess flux remains; this flux corresponds to the residual flux of the rest of the H\,{\sc i} tail, which shifts spatially to a southeastern direction toward HCG16-c (see Fig. \ref{fig:Panel}). The model parameters of the fitted Gaussian distribution are $\sigma$~=~24.2~$\pm$~2.8~km~s$^{-1}$ centered at 3552~$\pm$~2~km~s$^{-1}$.

In order to obtain a dynamic mass estimation, we used the approach from \cite{1996ApJS..105..269H}:

\begin{equation}
    M_{dyn} = 2.325\times10^5 \left( \frac{V_{rot}^2 + 3\sigma^2}{km^2\:s^{-2}}\right) \left(\frac{r}{kpc}\right)M_\odot.
\end{equation}

This equation is only valid in the case of dynamic equilibrium and self-gravitation. Some clues indicating that this could be the case here are the Gaussian shape of the H\,{\sc i} peak and the symmetry and sphericity of the optical counterpart, evidenced by the absence of residuals in the S\'ersic modeling, which contrasts with the complex morphology of similar galaxies in clear interaction \citep{2021MNRAS.tmp..419S}. However, a source of uncertainty is the low spectral resolution (21 km s$^{-1}$), which does not allow us to resolve dynamical substructures within the tip of the H\,{\sc i} tail. Taking these considerations into account, we proceeded to obtain this estimate. 

For the radius we used a range between a minimum value of two times the effective radius, which is 3 kpc, and a maximum value corresponding to the selected aperture in which the H\,{\sc i} spectrum was extracted, 40 arcsec and approximately 10 kpc at the group distance. These values produced a range of possible dynamic masses derived from the low resolution detection of H\,{\sc i}. Due to the apparent absence of a rotation pattern within the resolution limits \citep[see also][]{ 2019A&A...632A..78J}, we assumed that the dynamic mass is pressure supported, producing a value interval of M$_{dyn}$~=~[1.2, 4.1]~$\times$~10$^9$~M$_\odot$; this value is in contrast with the baryonic content, which is mostly gas, of M(HI)~=~3.9~$\times$~10$^8$~M$_\odot$. This would imply a mass-to-light ratio in the interval M/L~=~[3,~10]. It is worth commenting that even assuming a purely rotating unresolved component and a very conservative inclination of $i$~=~45$\degree$, the derived mass-to-light ratio would be in the range M/L~=~[1.5,~5]. Therefore, although there are caveats due to the low spectral resolution of the H\,{\sc i}, our analysis suggests that HCG 16-LSB1 contains a significant amount of dark matter; this is in contrast to what is expected for TDGs, which are typically poor in dark matter content \citep[e.g.,][]{ 2012ASSP...28..305D}. It is also worth commenting that even exclusively considering the observed baryonic matter of M(HI)~=~3.9~$\times$~10$^8$~M$_\odot$, it would be higher than the minimum mass for a long-lived TDG, which is approximately M~=~10$^8$~M$_\odot$ \citep{2006A&A...456..481B}. This would make the system gravitationally bound in future evolution, even assuming a total absence of dark matter. 

\begin{figure*}
    \centering
        \includegraphics[width=0.75\textwidth]{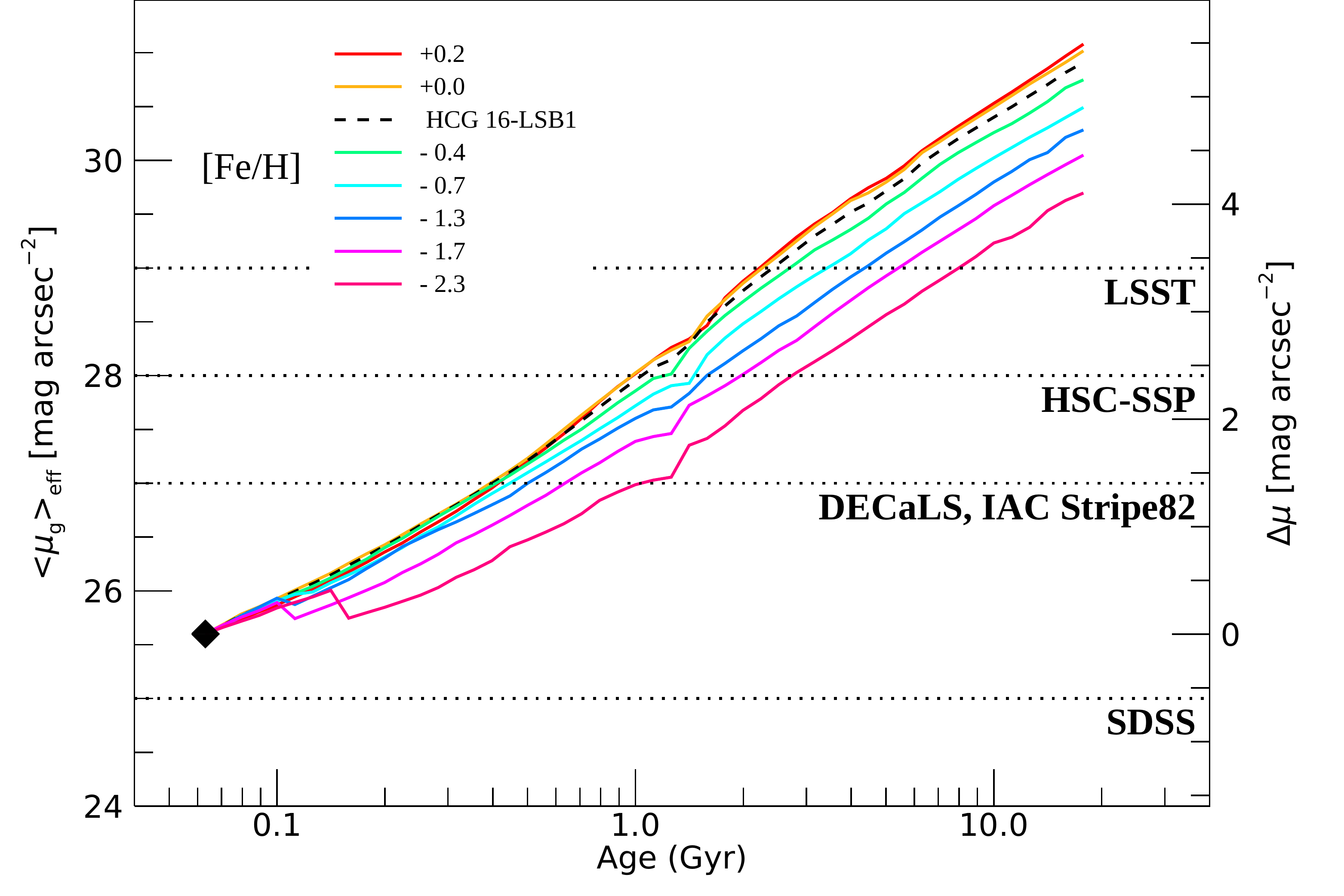}
    \caption{Evolution of the surface brightness of HCG 16-LSB1 in time. The black diamond symbol marks the observed age and surface brightness. The tracks in different colors mark the theoretical temporal evolution of the surface brightness that would correspond to different metallicities. The dashed black line corresponds to the evolution calculated for the metallicity of HCG 16-LSB1 ([Fe/H] = -0.16). Approximate limits on the maximum surface brightness of detectable galaxies are marked with dotted lines for the Sloan Digital Sky Survey \citep[SDSS; ][]{2000AJ....120.1579Y}, DECaLS \citep{2016MNRAS.460.1270D}, the IAC Stripe82 Legacy Survey \citep{2016MNRAS.456.1359F,2018RNAAS...2..144R}, the HSC-SSP \citep{2018PASJ...70S...4A}, and the future LSST \citep{2009arXiv0912.0201L}.}
    \label{fig:LSB_evo}
\end{figure*}

Although dynamic stability by self-gravitation seems assured, interactions with massive galaxies could easily tear the galaxy apart, which is often the case in compact groups. However, tidal features tend to eject at high speeds, being able not only to perform a distant orbit but even being able to flee the system by exceeding the escape velocity. HCG 16-LSB1 seems to have the latter ability. The escape velocity in HCG16 was estimated by \cite{2019A&A...632A..78J} to be 400~km~s$^{-1}$, close to the radial velocity of HCG 16-LSB1 with respect to the average of the host group, 300~km~s$^{-1}$. We can make a rough estimate of the tangential component of the velocity. Assuming (i) a flight time equal to the age of HCG 16-LSB1 and (ii) a distance traveled equal to half the projected distance between HCG 16-LSB1 and HCG16c (see Fig. \ref{fig:Panel}) -- considering its formation not in the vicinity of HCG16c but once the H\,{\sc i} tidal tail has already been ejected, which has been observed in other TDGs \citep{2014MNRAS.443.3601L} -- results in a tangential component of ${26\,kpc}/{58\,Myr}$ $\approx$ 450~km~s$^{-1}$. This is of course an estimate with large uncertainties, but it is useful to show that the tangential component of the velocity is significant. Therefore, the 3D component of the velocity is expected to be higher than the escape velocity of 400~km~s$^{-1}$, and therefore HCG 16-LSB1 will likely escape from the gravitational influence of the group and its interactions that could destroy it.

\subsection{Evolution in time of the optical surface brightness}

Since HCG 16-LSB1 will probably survive, it is interesting to explore how its photometric properties in the optical will evolve in the future. For that we can consider it as an SSP that is passively evolving, which is a good approximation since no current star formation is detected, the H\,{\sc i} peak is diffuse enough to not trigger new star formation \citep[2.45~$\times$~10$^{20}$~cm$^{-2}$; see][]{2019A&A...632A..78J}, and the accretion of new gas to the system is unlikely. For this we used the stellar population models from \cite{2015MNRAS.449.1177V} with a Kroupa universal IMF \citep{2001MNRAS.322..231K}. In Fig. \ref{fig:LSB_evo} we plot the photometric tracks corresponding to a temporal evolution, starting at the observed age and surface brightness of HCG 16-LSB1. We also plot photometric tracks for a wide range of metallicities, from [Fe/H] = $-$2.3 to $+$0.2, for comparison. Typical surface brightness limits for detecting low surface brightness galaxies are marked for the different surveys. We used as a reference the work by \cite{2018ApJ...857..104G}, who used the Hyper Suprime-Cam Subaru Strategic Program (HSC-SSP) with a detection limit for the mean surface brightness of approximately 28 mag arcsec$^{-2}$ in the \textit{g} band. We scaled this value using the limits derived from the Poisson noise through the relation given by \citet[][Appendix A]{2020A&A...644A..42R}, taking the various surveys' differing telescope apertures and exposure times into account. These calculated limits are an approximation since they not only depend on the depth of the data but also on the sizes of the sources and the method applied in the detection.

At the young age of HCG 16-LSB1, the stellar dimming due to the aging of the stellar populations is very strong. We can see that in just 400 Myr, the surface brightness will decrease enough to be undetectable in the optical data that we use in this work (DECaLS). In 1 Gyr of evolution, it would be undetectable in the deepest currently available survey, which is the HSC-SSP. And in 2 Gyr it would be undetectable in what will be the deepest optical survey in the near future, the Legacy Survey of Space and Time (LSST), reaching 29 mag arcsec$^{-2}$, which is the surface brightness of the faintest galaxy ever detected through integrated light \citep{2018ApJ...863L...7M}. After 10 Gyr of passive evolution, the stellar dimming is extreme, with a decrease in surface brightness of 5 mag arcsec$^{-2}$, making HCG 16-LSB1 completely undetectable, with an effective surface brightness above 30 mag arcsec$^{-2}$. It is interesting to note that the higher the metallicity, the stronger the dimming of the stellar populations. This means that in the case of TDGs, whose metallicity tends to be very high, this effect has a stronger impact on their detectability than that of ordinary dwarfs, whose metallicities are lower.

\section{Discussion}

The properties of HCG 16-LSB1 are peculiar. Its morphology and low surface brightness place it in the category of ultra diffuse galaxies {\citep[UDGs;][]{2015ApJ...798L..45V}}, that is, galaxies with large effective radii and low surface brightnesses {found in any environment \citep[e.g.,][]{2017MNRAS.468.4039R, 2017ApJ...842..133L, 2018MNRAS.481.4381M, 2019MNRAS.486..823R}. However,} its high metallicity and its location at the end of an H\,{\sc i} tail makes HCG 16-LSB1 compatible with a recently formed TDG. Strikingly, systematic detections of TDGs show compact-clumpy morphologies with bright surface brightnesses \citep[e.g.,][]{2020MNRAS.499.3399R}, contrary to that of HCG 16-LSB1. The dark matter content of HCG 16-LSB1 also appears anomalous for a TDG{, with a calculated mass-to-light ratio in the range M/L~=~[3, 10]. This is a standard for the general low-mass galactic population and would be in contradiction with the general characteristics of TDGs, which are generally poor in dark matter \citep[see e.g., ][]{2012ASSP...28..305D}. However, we warn that the spectral (and spatial) resolution that we have is not sufficient to obtain a reliable estimate of the dynamical mass of HCG 16-LSB1, dynamical calculations of such systems being extremely challenging in general \citep{2015A&A...584A.113L}. Future studies with better resolution will be necessary in order to confirm this significant presence of dark matter; however, if confirmed, this would make HCG 16-LSB1 an object of great interest, with {mixed} characteristics of both diffuse galaxies and TDGs.}  


Due to the low surface brightness of HCG 16-LSB1, we argue that its peculiar characteristics could be due to an observational bias in which the surface brightness limits of typical optical observations are insufficient to systematically detect such extreme objects. In fact, very deep optical observations focused on the study of TDGs find objects as faint as or even fainter than HCG 16-LSB1 \citep{2014MNRAS.440.1458D}. Therefore, it is expected that the new era of deep optical surveys will reveal a systematic presence of objects of this type, preferentially in strongly interacting galactic associations such as the Hickson Compact Groups.

Indeed, a correct understanding of observational biases is key to providing a correct representation of the low-mass galactic population. Our analysis, exemplified by the case of HCG 16-LSB1, shows that dimming via the passive evolution of stellar populations is another important factor to consider when studying galaxies without optical counterparts \citep[see also][]{ 2017ApJ...836..191T}. The characteristics that give rise to TDGs would be optimal in producing fake dark galaxies. In the first place, the ejection of H\,{\sc i} tails in which TDGs are formed would allow them to escape from the potential well of the host system, avoiding interactions that could destroy them. Additionally, since they are formed by reprocessed gas from massive host galaxies, TDGs are high in metallicity, maximizing the fading of their stellar populations, as clearly shown in Fig. \ref{fig:LSB_evo}. 

The connection between H\,{\sc i} debris and optically dark galaxies has been widely debated \citep[e.g.,][]{2005MNRAS.363L..21B, 2008ApJ...673..787D, 2017MNRAS.467.3648T}. While the fraction of H\,{\sc i} debris that actually contains a hidden stellar component is unknown, our work suggests that even {H\,{\sc i} debris galaxies} containing a detectable stellar component at the time of its formation, as is the case with TDGs, {are} able to reproduce the observational properties of optically dark galaxies observed in high-density environments \citep[e.g.,][]{2015AJ....149...72C, 2020A&A...642L..10B} after the passive evolution of their stellar content. This seems to be the case for HCG 16-LSB1. Our analysis shows that its most likely future involves its ejection out of the environment of HCG 16; it will appear as a gas cloud with a stellar population that will quickly fade beyond the optical detection limits. If the galaxy survives and is held together gravitationally -- even if it loses a significant fraction of its gas component, either by tidal interactions taking place in its formation or later on via photoevaporation -- its observational properties will be compatible with a dark matter halo in which only gas is detectable, with no optical counterpart, therefore mimicking the observational properties of dark galaxies without actually being one.

\begin{acknowledgements}
{We thank the anonymous referee for a careful review and useful comments}. We also thank Karen Lee-Waddell for helpful suggestions that improved this work. The authors acknowledge financial support from the grants AYA2015-65973-C3-1-R and RTI2018-096228-B-C31 (MINECO/FEDER, UE), as well as from the State Agency for Research of the Spanish MCIU through the “Center of Excellence Severo Ochoa” award to the Instituto de Astrofísica de Andalucía (SEV-2017-0709). JR acknowledges support from the State Research Agency (AEI-MCINN) of the Spanish Ministry of Science and Innovation under the grant "The structure and evolution of galaxies and their central regions" with reference PID2019-105602GB-I00/10.13039/501100011033. This research is based in part on observations made with the Galaxy Evolution Explorer, obtained from the MAST data archive at the Space Telescope Science Institute, which is operated by the Association of Universities for Research in Astronomy, Inc., under NASA contract NAS 5–26555, and is based in part on observations made with the Spitzer Space Telescope, which was operated by the Jet Propulsion Laboratory, California Institute of Technology under a contract with NASA.
\end{acknowledgements}

%
%

\begin{appendix}

\section{Reproducibility aspects}

This appendix summarizes the data sources and some reproducibility aspects. Our work was carried out using data from \cite{2019A&A...632A..78J} and public surveys. The optical data used in this work come from DECaLS \citep{2016MNRAS.460.1270D}, which is publicly accessible. The H\,{\sc i} data come from the work by \cite{2019A&A...632A..78J}, which includes a section on how to reproduce the full analysis: It provides Python Notebooks to produce the figures and a computational environment that includes the code and docker containers to run the analysis and produce the final data cube that we used here. The low resolution H\,{\sc i} cube used here can be found associated with the publication by \cite{2019A&A...632A..78J} and is publicly available; the additional high resolution H\,{\sc i} cube that we used has also been made publicly available under an open licence by \cite{Jones21}.

\end{appendix}

\end{document}